\documentclass[aps,prper,preprint]{revtex4-2}
\usepackage{amsmath}
\usepackage{amssymb}
\usepackage{graphicx}
\usepackage{xcolor}

\newcommand{\rev}[1]{#1}

\begin{document}

\title{A Dosage $\times$ Precision Framework for Instructional Effectiveness: Propositions, Boundary Conditions, and an Illustrative Case from Solid-State Physics}

\author{Kun Tao}
\email{taokun@lzu.edu.cn}
\author{ChengLong Jia}
\author{Desheng Xue}
\affiliation{Key Laboratory of Magnetism and Magnetic Functional Materials, Ministry of Education, Lanzhou University, Lanzhou, China}

\date{\today}

\begin{abstract}
We propose that instructional effectiveness may depend on an interaction between two design dimensions: dosage---the total duration of student-centered active learning allocated to a topic---and precision---the degree to which an intervention targets a specific, documented cognitive bottleneck with a mechanism-matched representational tool. This paper formalizes the framework, anchors it in cognitive load theory, derives its falsifiable predictions, and clarifies its incremental contribution relative to existing instructional-design literature. The central theoretical move is to decompose precision into three components---targeting ($P_t$), representational match ($P_r$), and scope restriction ($P_s$)---and to argue that only representational match acts directly on extraneous load. This decomposition yields a compact model in which the effective dosage is $D_{\mathrm{eff}}=D P_s$, the germane processing is $G=D P_s\eta(P_r)$ with $\eta$ the representational efficiency, and learning is $E=g(G)$ with $g$ increasing and concave \rev{(here $E$ is the cumulative learning outcome on the target construct, not a rate)}. The framework's core proposition (P3)---that precision can compensate for dosage---then becomes a quantifiable substitutability claim, $\mathrm{d}D/\mathrm{d}P_r=-\eta/\eta'$, rather than a mere existence statement; its scope is bounded by a precision ceiling and a minimum dosage threshold. We distinguish the framework from Carroll's classic time-needed model by the directional character of precision and by the estimability of the substitution rate, and from the ICAP framework by distinguishing ``matching of participation content to the bottleneck'' from ``mode of participation.'' As an illustrative case (not a test of the framework), we report a 15-minute three-dimensional visualization intervention in a required solid-state physics course ($N=82$), engineered to target a single documented bottleneck (mental rotation of reciprocal space). Gains concentrated on the core-bottleneck items ($+27.1$ pp), while the combined gain on near-transfer and far-transfer items was substantially smaller ($+4.9$ pp)---a pattern consistent with the framework's prediction. The case illustrates rather than tests the framework; the substitutability claim (P3) remains a falsifiable prediction for future multi-condition studies.
\end{abstract}

\maketitle

\section{INTRODUCTION}

How should effective instructional interventions be designed when class time is limited? This is a common problem faced by instructors across disciplines. The literature has extensively documented the overall effectiveness of active learning [1,2], but there is still a lack of an actionable framework for how two design dimensions interact: ``how much time is invested in given content'' (dosage) and ``what that time is aimed at'' (precision).

Cognitive load theory [3,4] provides part of the foundation for this problem. It distinguishes intrinsic load, extraneous load, and germane load, and predicts that reducing extraneous load frees working memory for germane processing. However, cognitive load theory itself does not distinguish ``dosage'' and ``precision'' as independent design dimensions, nor does it predict their interaction. In the existing instructional-design literature, Merrill's ``First Principles of Instruction'' [5] emphasizes problem-centeredness and activation of prior knowledge, but does not discuss the trade-off between dosage and focus; van Merri\"enboer et al.'s ``Four-Component Instructional Design'' [6] discusses the allocation of learning tasks and supportive information, but does not formalize the substitutability of time and precision. Rosenshine's principles of instruction [7] emphasize explicit teaching and small-step guidance, whose core intuition is ``focus teaching on where students are most likely to go wrong''---which is highly relevant to the targeting component ($P_t$) of the present framework---but Rosenshine provides practical principles and does not formalize them into quantifiable design dimensions. Recent studies on visualization-based teaching in solid-state physics [8,9] and work integrating computational simulations into undergraduate teaching [10--13] further suggest that the effectiveness of an instructional tool may depend on how specifically it targets a particular learning difficulty, but these studies do not formalize this idea into testable design dimensions.

The closest antecedent is Carroll's model of school learning [14], in which the degree of learning is a function of the ratio of time actually spent to time needed, and in which instructional quality acts by reducing the time needed. This framework shares Carroll's core intuition that time is not the only lever, and we therefore state the relation explicitly (Sec.~II.D). In brief, Carroll's ``quality'' is undirected---it is not indexed to a specific cognitive bottleneck---whereas precision in the present framework is directional (a property of the match between a representation and an identified bottleneck) and yields an estimable substitution rate rather than a ratio. Bloom's analysis of individual tutoring [15] is likewise the classic demonstration that highly targeted instruction can vastly outperform whole-class instruction, and the present framework can be read as an attempt to make that observation quantitative and design-actionable.

\emph{Distinction from the ICAP framework.} The ICAP framework [16] concerns \emph{how} students participate (mode of engagement: passive, active, constructive, interactive) and predicts that interactive engagement outperforms passive reception. The present framework is orthogonal to ICAP: ICAP does not predict \emph{what the participation content should match} in terms of cognitive bottleneck. An interactive group discussion whose content deviates from the bottleneck (low $P_t$), or that requires students to perform the bottleneck transformation themselves rather than externalizing it (low $P_r$), would still count as a higher-order mode of engagement in ICAP, but would be a low-precision intervention in the present framework. Conversely, a well-designed demonstration (which in ICAP might count only as ``active'') that directly externalizes the bottleneck (high $P_r$) and strictly restricts its scope (high $P_s$) would be a high-precision intervention in the present framework. ICAP therefore explains the effect hierarchy of ``mode of participation,'' while the present framework explains how the match between ``participation content'' and the bottleneck moderates that hierarchy. The two are complementary rather than competing.

\emph{Terminological caution.} ``Precision'' here is unrelated to the ``Precision Teaching'' tradition in behavior analysis [17], which refers to frequent measurement and charting of response fluency. The two uses of the word share no content, and we flag the collision explicitly to prevent misreading.

The direct contribution of this paper is to propose and formalize a dosage--precision framework describing the interaction of these two dimensions, derive falsifiable predictions, and anchor the framework in cognitive load theory. The incremental contribution is not to rediscover cognitive load theory, but to derive from it a new, quantifiable prediction about the interaction of design variables (P3): under specified conditions, precision and dosage can compensate for each other, and the degree of substitutability is estimable.

As an illustrative case for the framework, we report an intervention deliberately set at a minimal dosage: a 15-minute three-dimensional visualization demonstration in a required solid-state physics course. We emphasize that the function of this case is to illustrate the framework rather than to test it---it shows that the pattern predicted by the framework can indeed be observed in a low-dosage $\times$ high-precision region, and it provides a replicable starting point for future comparative studies that genuinely test the framework. The present study contains one condition only; no control or comparison arm was available under the curricular constraints in which the data were collected, and we therefore make no causal claim about precision.

This study has three goals: (1) to propose the dosage--precision framework and formalize its core propositions; (2) to anchor the framework in cognitive load theory, derive its falsifiable predictions, and clarify its incremental contribution relative to existing literature; and (3) to illustrate, through a concrete case, how the framework can be applied in an actual instructional setting.

\section{THEORETICAL FRAMEWORK: THE DOSAGE $\times$ PRECISION INTERACTION}

\subsection{Definitions of the two dimensions}

We propose that instructional effectiveness is not determined by a single lever, but by the interaction of two design variables:

\begin{itemize}
\item \textbf{Dosage ($D$)}: the total duration of student-centered active learning allocated to the target content. We note that duration is only one facet of what is colloquially called ``time on task'': the literature on instructional time distinguishes allocated time, engaged time, and academic learning time [18,19], and the ICAP framework distinguishes passive, active, constructive, and interactive engagement at equal duration [16]. In this paper we take $D$ to denote allocated duration and treat engagement quality as a separate factor held fixed by design; we return to this in the boundary conditions (Sec.~IV.B).
\item \textbf{Precision ($P$)}: the degree to which an intervention is aligned with a specific, documented cognitive bottleneck.
\end{itemize}

Effectiveness, $E$ \rev{(the cumulative learning outcome on the target construct)}, is proposed to be a function of the pair $(D,P)$. \rev{We emphasize that $E$ denotes a cumulative outcome, not a rate: doubling the dosage in the linear regime roughly doubles $E$ rather than leaving it unchanged. Where a marginal (per-minute) quantity is intended, we write the derivative $\partial E/\partial D$ explicitly.}

\rev{Accordingly, the dosage threshold $D^{*}$ introduced below is defined as the point at which the \emph{total} learning outcome saturates---beyond it, additional dosage no longer increases $E$---and not as a point of declining efficiency.}

\emph{A three-component decomposition of precision.} A single scalar $P$ is not sufficient, because three logically distinct properties are conflated in the phrase ``targeted instruction.'' We therefore decompose
\begin{equation}
P = (P_t,\,P_r,\,P_s),
\end{equation}
with the following meanings and roles.

\begin{enumerate}
\item \textbf{Targeting ($P_t$; binary).} Whether the intervention addresses a cognitive bottleneck that has been documented in the literature. $P_t$ is a precondition for the framework rather than a term in the effectiveness equation: if $P_t=0$, students may still learn, but they are not learning the intended construct, and gains will not appear on instruments measuring that construct. We therefore regard $P_t=1$ as an inclusion criterion for applying the framework, not as a manipulated variable. We treat it as binary for simplicity; in principle it could be continuous (the degree to which the intervention's content maps onto the documented bottleneck), but we do not develop that extension here.
\item \textbf{Representational match ($P_r$; continuous, $[0,1]$).} The degree to which the intervention's representational format eliminates the specific transformation that constitutes the bottleneck. For the reciprocal-space bottleneck, that transformation is mental rotation. $P_r$ is the only component that acts directly on extraneous load, and it is therefore the component that drives the interaction between dosage and precision.
\item \textbf{Scope restriction ($P_s$; continuous, $[0,1]$).} The degree to which the intervention's content is restricted to the bottleneck and its immediate conceptual consequences. Content that wanders beyond this scope dilutes the intervention: at fixed clock time, a broader scope means less time on the target. $P_s$ therefore acts on the effective dosage rather than on load.
\end{enumerate}

This decomposition matters operationally. An intervention that uses an externalizing 3D representation (high $P_r$) but also reviews unrelated material (low $P_s$) is not simply ``high-precision''; it is high on one component and low on another, and the framework predicts correspondingly different outcomes. Likewise, a well-scoped intervention that nonetheless requires students to perform the bottleneck transformation themselves (high $P_s$, low $P_r$) should not be expected to show the substitutability predicted by P3.

\subsection{A formal model and its propositions}

Let $\eta(P_r)\in[0,1]$ denote the \emph{representational efficiency}: the fraction of effective instructional time that is converted into germane processing rather than consumed by decoding the representation itself. We assume $\eta$ is strictly increasing and concave, with $\eta(0)=\eta_0>0$ and $\eta\to\eta_{\max}\le 1$ as $P_r\to 1$. The ceiling $\eta_{\max}$ corresponds to the point at which extraneous load from the representation can no longer be reduced; the remaining load is then intrinsic load, determined by the inherent complexity of the concept and not removable by instructional design [3]. We refer to it as the \emph{precision ceiling}. Define the effective dosage and the germane processing as
\begin{equation}
D_{\mathrm{eff}} = D P_s, \qquad G = D_{\mathrm{eff}}\,\eta(P_r) = D P_s\,\eta(P_r),
\end{equation}
and let the learning outcome be
\begin{equation}
E = g(G),
\end{equation}
with $g$ strictly increasing and concave. The quantity $G$ is the amount of germane processing that is actually directed at the target construct. \rev{As stated in Sec.~II.A, $E$ is a cumulative outcome; the marginal return of dosage is $\partial E/\partial D$, which is a rate and is used as such in P1 below.}

\emph{Justification of the concavity assumptions.} The two concavity assumptions are the mathematical cornerstone of P1 and P3, and their plausibility requires justification.

\emph{Concavity of $g$.} Learning curves in cognitive science are typically described by a power-law form, $E\propto G^{\alpha}$ with $0<\alpha<1$ [20], \rev{or by an exponential form [21]; the empirical dispute over which functional form is preferred does not affect the present argument, because both are strictly increasing and concave.} A power-law (or exponential) function is strictly increasing and concave. In conceptual-learning settings, the concavity assumption means that, once the bottleneck-specific schema has been constructed, the marginal benefit of additional germane processing diminishes. The framework's scope is limited to settings where the bottleneck has been identified and students possess the necessary prior knowledge---in which the concavity approximation holds. For novices with substantial gaps in prior knowledge, the learning curve may be $S$-shaped, in which case P1 holds only after $G$ exceeds the inflection point; this qualification is discussed under boundary conditions (Sec.~IV.B).

\emph{Concavity of $\eta$.} The concavity assumption on $\eta$ means that the marginal benefit of representational match diminishes: moving from low $P_r$ to moderate $P_r$ eliminates a large amount of extraneous load, whereas further improvement from high $P_r$ to higher $P_r$ yields limited benefit. This assumption holds naturally as $\eta$ approaches its ceiling $\eta_{\max}$. It must be acknowledged that for some interventions (e.g., the transition from complete absence of externalization to complete externalization), $\eta$ may be closer to a piecewise-concave or step function; but within the range of interest to the framework (continuous variation of $P_r$ from low to high), concavity is a reasonable approximation.

\emph{Sensitivity.} If $g$ is $S$-shaped rather than concave, the substitution rate of P3 remains negative after $G$ exceeds the inflection point (precision can still compensate for dosage), but may fail below the inflection point (because there the marginal benefit of additional germane processing is increasing). If $\eta$ is linear rather than concave, the substitution rate $-\eta/\eta'$ changes in numerical value but remains negative---the substitution relation between precision and dosage is qualitatively preserved. The qualitative prediction of P3 is therefore robust to the choice of functional form; the precise value of the quantitative substitution rate depends on the specific functional form and must be estimated from data in future research.

Three propositions follow.

\begin{enumerate}
\item \textbf{P1 (diminishing returns of dosage).} Because $g$ is concave, $\partial E/\partial D = P_s\eta(P_r)g'(G)$ is decreasing in $D$; there exists a dosage threshold $D^{*}$ beyond which the marginal return of additional dosage approaches zero. \rev{Since $E$ is a cumulative outcome (Sec.~II.A), $D^{*}$ marks the saturation of the total learning outcome: for $D>D^{*}$, further dosage leaves $E$ essentially unchanged.} The threshold is reached sooner (at smaller $D$) the larger $\eta(P_r)$, i.e., high-precision interventions saturate earlier.
\item \textbf{P2 (precision matters most when dosage is low).} Holding $D$ fixed,
\begin{equation}
\frac{\partial E}{\partial P_r} = g'(G)\,D P_s\,\eta'(P_r).
\end{equation}
At low $D$, $G$ is small, so $g'(G)$ is large (by concavity of $g$); hence the marginal benefit of representational match is larger at low dosage than at high dosage. This is the formal interaction: $P_r$ and $D$ are substitutes, not merely additive contributors.
\item \textbf{P3 (precision can compensate for dosage).} Consider the level set $G=\text{const}$, i.e., the combinations of dosage and representational match that yield the same germane processing. Along this set, taking the total differential of $G=D P_s\eta(P_r)=\text{const}$ gives $P_s\eta\,\mathrm{d}D + D P_s\eta'\,\mathrm{d}P_r = 0$, which rearranges to
\begin{equation}
\frac{\mathrm{d}D}{\mathrm{d}P_r} = -\frac{\eta(P_r)}{\eta'(P_r)} < 0.
\end{equation}
The magnitude of this derivative is the \emph{substitution rate}: it states how many minutes of dosage are required to compensate for a unit loss of representational match, and vice versa. P3 is therefore not an existence claim (``there exist two points such that \ldots'') but a quantitative statement about a derivable, estimable quantity.
\end{enumerate}

\emph{Boundary conditions.} P3 is not universal. Two boundaries delimit its applicability. First, the \emph{precision ceiling}: as $P_r\to 1$, $\eta'\to 0$, so the substitution rate $\mathrm{d}D/\mathrm{d}P_r\to-\infty$; precision can no longer buy dosage once extraneous load is near zero. Second, the \emph{minimum dosage threshold} $D_{\min}$: if $D$ is so small that $G$ falls below the level required for basic schema construction, then $E\to 0$ regardless of $P_r$. Combining both, the precise formulation is: for $D>D_{\min}$ and $P_r<P_{r,\max}$, precision and dosage are substitutable with rate $-\eta/\eta'$; outside this region the compensatory relation does not hold.

\subsection{Mechanism: anchoring in cognitive load theory}

The model is grounded in cognitive load theory [3,4]. Learning occurs when working-memory resources are occupied by germane processing rather than exhausted by extraneous load. The three components of precision map onto distinct constructs.

\emph{Representational match acts on extraneous load.} A representation that eliminates the bottleneck transformation removes the associated extraneous load, raising $\eta$. This is the mechanism behind P2 and P3: at low $D$, a short time budget is either mostly consumed by decoding the representation (low $P_r$, low $\eta$) or mostly converted into germane processing (high $P_r$, high $\eta$).

\emph{Scope restriction acts on effective dosage.} Content outside the bottleneck does not reduce load; it divides the time budget. This is why $P_s$ enters multiplicatively with $D$ rather than through $\eta$.

\emph{Targeting is a precondition.} $P_t$ ensures that the germane processing produced is directed at the construct of interest.

Two qualifications from the cognitive-load literature must be stated explicitly. First, the tripartite intrinsic/extraneous/germane division used above is the original formulation; the status of ``germane load'' as a separate additive source has since been revised, and we use the term here descriptively (to denote processing directed at schema construction) rather than as a third additive reservoir [4]. Second, reducing extraneous load is not unconditionally desirable: some difficulty is \emph{desirable} because it supports transfer and retention [22], and generating solutions before instruction can improve later learning [23]. The framework therefore concerns specifically \emph{extraneous} load arising from representational mismatch; it does not claim that all difficulty should be minimized. An intervention that removes germane difficulty in the name of ``precision'' would be predicted to improve immediate performance while impairing transfer---a prediction we return to below.

\subsection{Incremental contribution: dialogue with Carroll's model}

Carroll's model of school learning [14] holds that the degree of learning is determined by the ratio of time spent to time needed, with instructional quality acting to reduce the time needed. The present framework is clearly in this lineage, and we state the relation precisely.

Carroll's ``quality of instruction'' is undirected: it is a global property of instruction, not indexed to any particular cognitive operation, and the model does not distinguish among the several ways in which instruction can be improved. The dosage--precision framework makes three additions.

\begin{enumerate}
\item \textbf{Directionality.} Precision is defined relative to an identified bottleneck and a specific representational transformation (Eq.~1). Two interventions of equal global ``quality'' may differ in $P_r$ if one matches and the other mismatches the bottleneck. This is not expressible in Carroll's quality term.
\item \textbf{Separability and heterogeneity.} The decomposition separates targeting, representational match, and scope. Carroll's model treats quality as one quantity; the framework predicts that $P_r$ and $P_s$ have different effects (on load vs.\ on effective dosage) and therefore different empirical signatures.
\item \textbf{Estimability.} The framework yields an explicit substitution rate, Eq.~(5), which is in principle measurable by mapping an iso-$G$ contour. Carroll's ratio is a comparative statement, not an estimable derivative.
\end{enumerate}

A second body of work, Merrill's first principles [5] and the 4C/ID model [6], contains the idea that instruction should target learners' difficulties, but neither separates ``what is targeted'' from ``how much time is invested'' as independently variable dimensions, and neither derives an interaction prediction. Rosenshine [7] provides practical principles for focused instruction but does not formalize them. The framework's contribution is thus to take an intuition present in several traditions---that focused instruction is efficient---and render it as a two-dimensional model with a quantified, falsifiable substitutability claim.

\subsection{How the model can be falsified}

The framework makes five predictions, ordered from lower to higher testing difficulty. Table~\ref{tab:predictions} summarizes them and their status in the illustrative case.

\begin{enumerate}
\item \textbf{Differential-precision pattern.} Gains are predicted to be largest on items measuring the targeted bottleneck and on misconceptions embedded in it, and smaller on non-targeted content. We emphasize that this pattern \emph{cannot by itself establish} that $P_r$ is the cause of the concentration: since a high-$P_t$, high-$P_s$ intervention directs its whole content at the target, some concentration follows from the definition of the intervention alone. A direct test requires varying $P_r$ at fixed content, dosage, and assessment.
\item \textbf{Conservation of time.} If the intervention replaces rather than adds to existing instruction, any gain is a gain in efficiency, which is the premise underlying P3. A direct test compares substitutive and non-substitutive designs.
\item \textbf{Dose--response curvature.} At high $P_r$, additional dosage beyond the saturation threshold yields little further gain (P1); at low $P_r$, the same additional dosage yields larger absolute gains. Testing requires at least two dosage levels at fixed $P_r$, and ideally both precision levels.
\item \textbf{Mechanism transfer.} Identifying the bottleneck in a different concept and designing a matched intervention should reproduce the differential-precision pattern. Testing requires repeating the design in another conceptual domain.
\item \textbf{Bottleneck-nature moderation.} This is the most specific prediction generated by the formalization. If $\eta$ reflects representational efficiency, then the quantity governing the precision advantage is the relative slope $\eta'(P_r)/\eta(P_r)$ (from Eq.~5), and this ratio should be larger for bottlenecks whose difficulty is more representational, i.e., bottlenecks that depend more heavily on mental transformation of spatial, graphical, or symbolic structures. Equivalently: the more representational the bottleneck, the larger the marginal benefit of precision, and the larger the substitution rate of precision for dosage.
\end{enumerate}

\emph{Operationalizing Prediction 5.} The representational nature of a bottleneck can be assessed by (i) expert task analysis, in which two independent raters score, on a 5-point scale, the extent to which the bottleneck requires spatial--representational processing (three-dimensional rotation, graphical transformation, symbolic manipulation), with inter-rater agreement reported; and (ii) correlation of bottleneck difficulty with established spatial-ability measures (e.g., mental-rotation tests [24,25]). A test then compares two bottlenecks---one high-, one low-representational---under the same dosage and a comparable precision manipulation, with the precision comparability ensured by verifying that both interventions satisfy the three precision criteria. If the precision effect is significantly larger for the high-representational bottleneck, Prediction 5 is supported; if the two do not differ, the mechanistic account (that precision acts on representational extraneous load) is challenged. Under the formalization, Prediction 5 is no longer an independent fifth claim: it is a statement about the shape of $\eta(P_r)$, and it is therefore the framework's most falsifiable component.

\emph{Estimating $\eta(P_r)$.} The substitution rate in Eq.~(5) is estimable only if $\eta(P_r)$ can be recovered from data. This requires at least two conditions that differ in $P_r$ at fixed content, dosage, and $P_s$. Under the assumption that $g$ is common across conditions, the ratio of learning gains identifies the ratio $\eta(P_{r1})/\eta(P_{r2})$; if $g$ is further assumed to be a power function $g(G)=G^{\alpha}$, then
\begin{equation}
\frac{\eta(P_{r,\text{high}})}{\eta(P_{r,\text{low}})} = \left(\frac{E_{\text{high}}}{E_{\text{low}}}\right)^{1/\alpha}.
\end{equation}
For example, if $\alpha=0.5$, a twofold difference in learning gains implies a fourfold difference in efficiency. The present study does not attempt this estimation; it provides one point only. We state the estimation strategy here so that the framework's quantitative claim is not merely formal.

\rev{\emph{Identification condition.} Equation (6) is not identified from a single pair of conditions. One gain ratio provides one equation, but it contains two unknowns---the efficiency ratio $\eta(P_{r,\text{high}})/\eta(P_{r,\text{low}})$ and the exponent $\alpha$. The system is therefore underdetermined, and the numerical example above presupposes a fixed value of $\alpha$. Estimation requires either (i) fixing $\alpha$ from independent practice-curve estimates in the literature, or (ii) adding at least a third precision level at fixed content and dosage, so that $\alpha$ and the shape of $\eta(P_r)$ can be fitted jointly. We state this condition explicitly to avoid overstating the estimability of Eq.~(6); without it, the expression should be read as a definition of the required measurement design rather than as a ready estimator.}

\emph{Operationalizing $P_r$.} In the present case $P_r$ is a design judgment and is not assigned a numerical value. Future comparative studies testing the framework can operationalize $P_r$ through the following complementary strategies: (1) \emph{expert blind rating}: invite solid-state physics education experts unfamiliar with the present framework to rate, on a 5-point scale, the extent to which the intervention's representation ``eliminates the need for students' mental rotation,'' with inter-rater agreement (e.g., ICC) reported as validity evidence for $P_r$; (2) \emph{student cognitive-load self-report}: use NASA-TLX or a cognitive-load scale to measure the mental demand students experience in interpreting the representation; high-$P_r$ interventions should report lower extraneous-load ratings; (3) \emph{process data}: in digital environments, measure the time students take from presentation of the representation to correctly answering the bottleneck items; higher $P_r$ should yield shorter times (controlling for prior knowledge). Future comparative studies testing the framework should adopt at least strategies (1) and (2).

\rev{Strategy (1) is the cheapest to implement and requires no new student data. We therefore report it as the minimum operationalization that any future test of the framework should include, and we return to it in Sec.~IV.C as a concrete next step for the present case.}

\begin{table}[htbp]
\caption{Predictions of the dosage $\times$ precision framework and the patterns observed in the illustrative case reported in this paper. ``Consistent with prediction'' means the observed pattern matches the prediction but does not establish the proposed mechanism as the cause. ``Direct test'' requires varying precision at fixed content and dosage (Prediction 1), comparing substitutive and non-substitutive designs (Prediction 2), multiple dosage levels (Prediction 3), a different conceptual domain (Prediction 4), or a cross-bottleneck comparison (Prediction 5); all lie beyond the design scope of the present case.}
\label{tab:predictions}
\begin{ruledtabular}
\begin{tabular}{p{4.2cm}p{5.4cm}p{3.4cm}}
Prediction & Pattern in this case & Status\\
\hline
1. Differential-precision pattern & core-bottleneck items $+27.1$ pp; targeted misconception $65.9\%\to6.1\%$; near/far-transfer items combined $+4.9$ pp & Consistent with prediction; not a direct test\\
2. Conservation of time & 15-min demonstration replaces standard segment; large retained gain & Consistent with prediction; direct test requires substitutive vs.\ non-substitutive design\\
3. Dose--response curvature (P1) & Not applicable (single dosage level) & Future research\\
4. Mechanism transfer & Not applicable (single concept) & Future research\\
5. Bottleneck-nature moderation & Not applicable (single bottleneck) & Future research\\
P3 substitutability & Gains observed only in the low-$D\times$high-$P$ region; no high-$D\times$low-$P$ arm & Not tested; future research\\
\end{tabular}
\end{ruledtabular}
\end{table}

\section{ILLUSTRATIVE CASE: A 15-MINUTE THREE-DIMENSIONAL VISUALIZATION INTERVENTION}

\emph{Note.} The function of the following case is to illustrate how the dosage--precision framework can be applied to a concrete instructional setting and to show that the pattern it predicts can be observed in a low-dosage $\times$ high-precision region. It does not constitute a test of the framework's propositions: the case contains one condition (low $D\times$high $P$), lacks the high-$D\times$low-$P$ condition required for comparing P3, and includes no manipulation of $P_r$ at fixed content. Detailed implementation guidelines (minute-by-minute script, VESTA operation steps, complete test forms, DFT parameters) are provided in the Supplemental Material [31].

\subsection{Case design}

The case is a ``computation--visualization--cognition'' (C-V-C) teaching progression consisting of three conceptual layers, informed by Collins' cognitive apprenticeship theory [26]:

\begin{enumerate}
\item \textbf{Computation layer} (instructor preparation, zero student time): the instructor performs all DFT calculations in advance using VASP [27], and output files (CHGCAR, DOSCAR, EIGENVAL) are distributed to students. Students do not touch any computational tools.

\item \textbf{Visualization layer} (student activity): students open the pre-computed files in VESTA [28] (Visualization for Electronic and Structural Analysis, a free, cross-platform three-dimensional visualization program) and, through point-and-click operations, rotate, zoom, and inspect three-dimensional crystal structures and Brillouin zones, as shown in the Fig.\ref{fig1}.

\begin{figure}[htbp]
  \centering
\center\includegraphics[width=12 cm,clip=true]{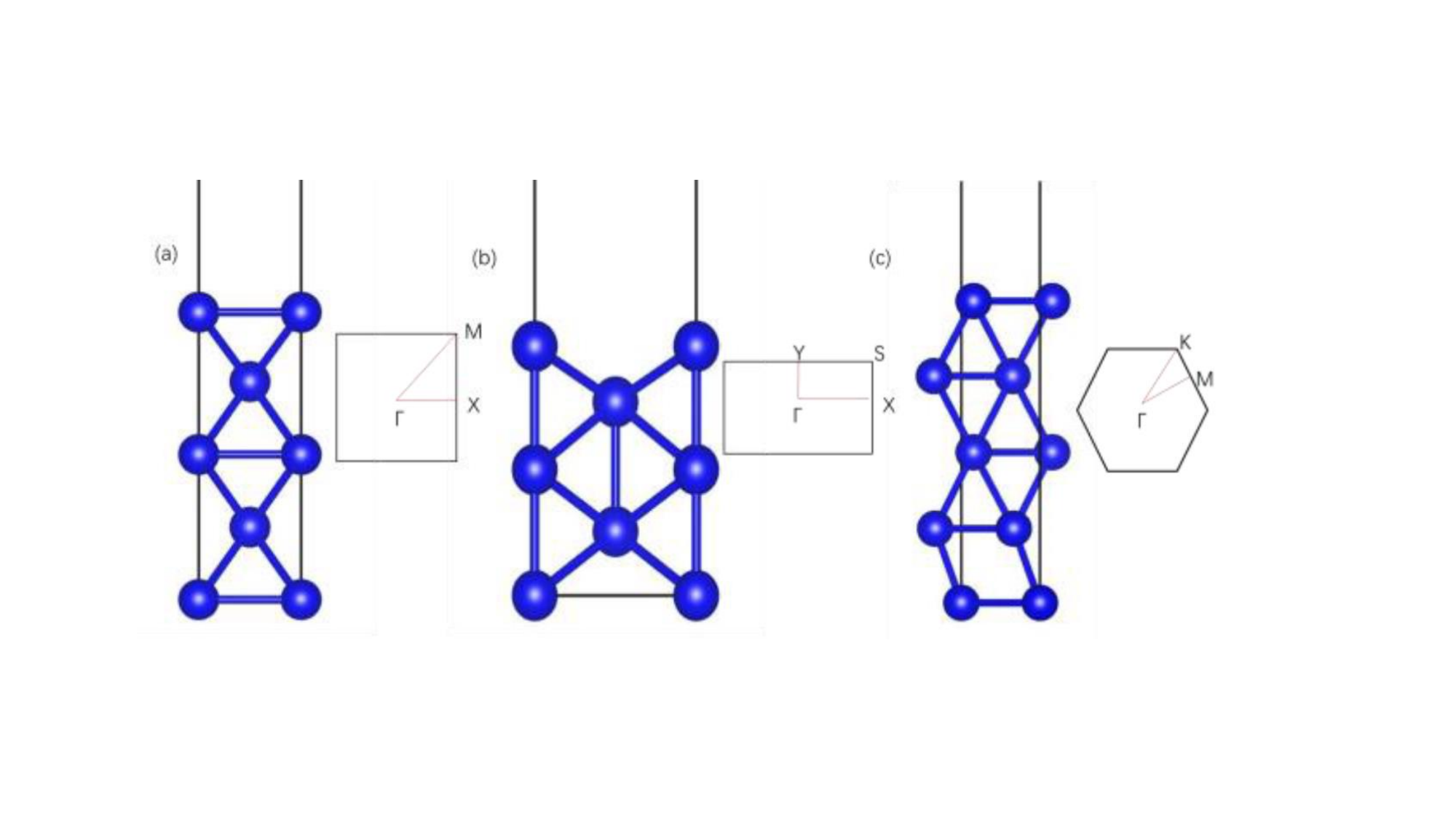}
\caption{\label{fig1}
Three surface structures of Ni and their corresponding two-dimensional Brillouin zones. (a) 001, (b) 110, (c) 111.}
\end{figure}

\item \textbf{Cognition layer} (whole-class discussion): the instructor leads a structured discussion that confronts a documented misconception---``higher DOS near $E_F$ implies greater stability''---using pre-computed d-band center values, as shown in the Fig.\ref{fig2}.
\end{enumerate}

\begin{figure}[htbp]
  \centering
\center\includegraphics[width=12 cm,clip=true]{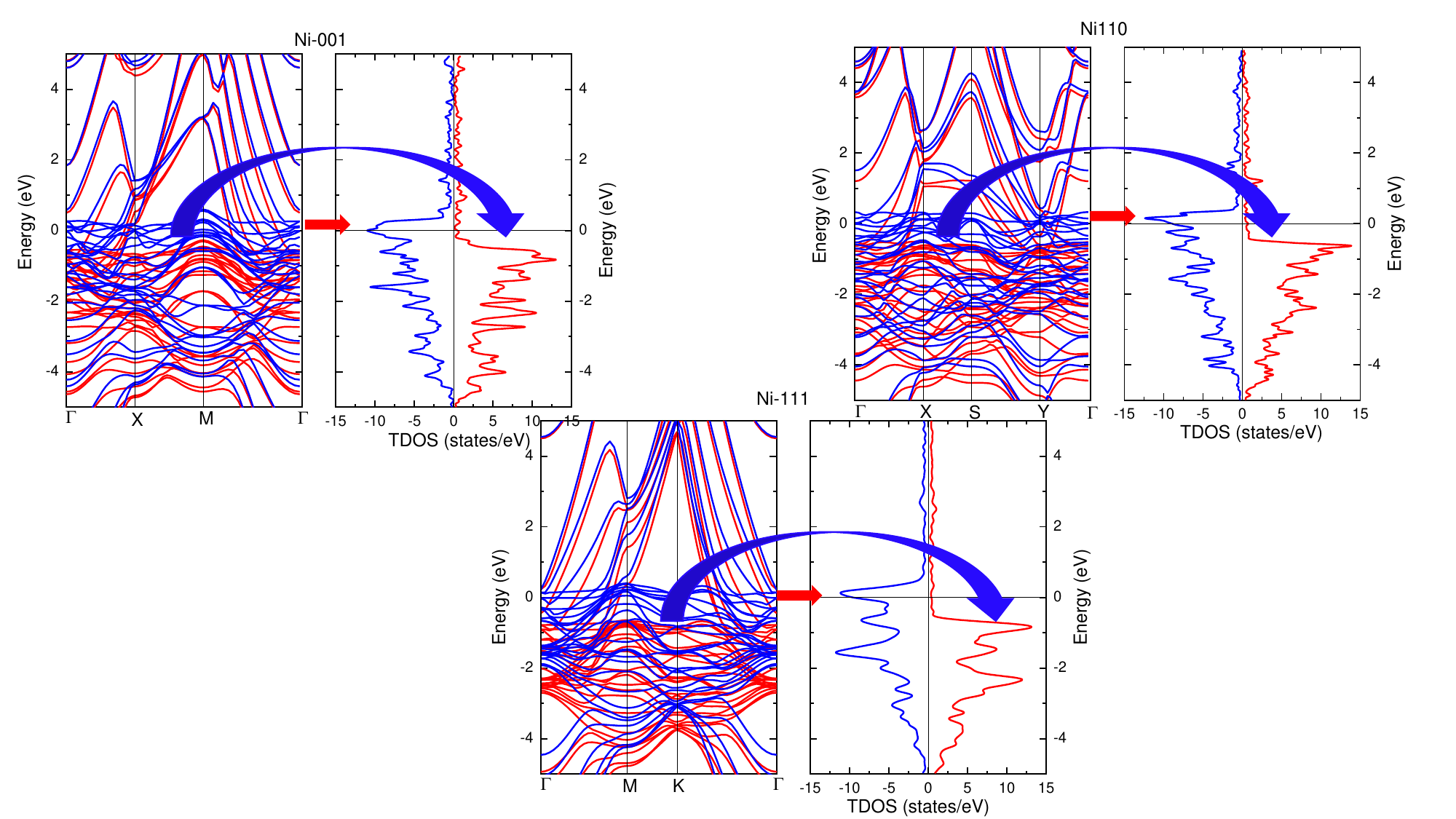}
\caption{\label{fig2}
Band structures and corresponding total density of states of the three Ni low-index surfaces: (a) (001), (b) (110), (c) (111). Insets show the high-symmetry points within the respective two-dimensional Brillouin zones.}
\end{figure}

According to the decompositions of Sec.~II, this intervention is characterized as follows. \emph{Dosage}: $D=15$ minutes, replacing rather than adding to a standard lecture segment. \emph{Targeting}: $P_t=1$; the bottleneck (mental rotation and three-dimensional reconstruction of reciprocal space) is documented in the spatial-ability literature [24,25]. \emph{Representational match}: $P_r$ high; the tool directly externalizes the three-dimensional structure that students would otherwise have to reconstruct mentally. \emph{Scope restriction}: $P_s$ high; the content is confined to the bottleneck and its immediate consequences. Note that $P_r$ here is a design judgment, not a measured quantity; no manipulation check was performed, and we therefore cannot assign it a numerical value.

\subsection{Participants and assessment}

Eighty-two third-year physics majors (two consecutive cohorts, $n=40$ and $n=42$) participated. The assessment instrument consisted of 10 multiple-choice items measuring three constructs: Brillouin-zone recognition and spatial reasoning (4 items), high-symmetry point identification (3 items), and band--DOS relationship reasoning (3 items). All items were objective multiple-choice with a single correct answer, so scoring involved no rater judgment and therefore no inter-rater reliability issue. Content validity was established in three steps: item--objective mapping, distractor design based on documented student difficulties [25,29], and review by two experts. The posttest was a parallel form administered five weeks after the intervention and judged by the two experts to be slightly more difficult than the pretest. Item-level response data were not retained (the students have graduated; the archival record contains only total scores), so internal-consistency statistics such as Cronbach's $\alpha$ could not be computed; both complete forms are provided in the Supplemental Material so that these properties can be established in replication studies. Classroom feedback consists of the instructor's contemporaneous formative notes, not a systematically coded dataset, and has no inter-rater reliability; it is reported only as illustrative examples.

\emph{Item classification framework.} Different steps of the intervention cover the three types of items to different degrees, so we classify the 10 items by their distance from the bottleneck (mental rotation and three-dimensional reconstruction of reciprocal space) into three categories:

\begin{itemize}
\item \textbf{Core-bottleneck items (4 items)}: directly measure mental rotation and three-dimensional reconstruction (e.g., identifying the Brillouin-zone shape for different surface orientations). Steps 1 and 2 of the intervention directly target the abilities these items measure.
\item \textbf{Near-transfer items (3 items, high-symmetry point identification)}: concern the same conceptual domain (reciprocal space) but do not directly require mental rotation. In the intervention, these points ($\Gamma$, X, Y, M, K) were labeled in VESTA; students may have acquired procedural familiarity rather than structural understanding.
\item \textbf{Far-transfer items (3 items, band--DOS reasoning)}: concern a different conceptual domain (electronic structure) and are only indirectly involved in Step 3 and the cognition-layer discussion.
\end{itemize}

This classification is more precise than a simple ``target/non-target'' dichotomy, and reflects the different degrees to which the intervention covers the different items. Because item-level response records were not retained, near-transfer and far-transfer items cannot be reported separately; their combined gain is $+4.9$ pp.

\subsection{Core results}

\begin{table}[htbp]
\caption{Core results of the illustrative case.}
\label{tab:results}
\begin{ruledtabular}
\begin{tabular}{lccc}
Measure & Pretest & Posttest & Gain\\
\hline
Total conceptual score (10 items) & 49.5\% & 63.3\% & $+13.8$ pp ($d_z=1.73$)\\
Core-bottleneck items (4 items) & 54.6\% & 81.7\% & $+27.1$ pp\\
Near-transfer $+$ far-transfer items (6 items, combined) & 46.1\% & 51.0\% & $+4.9$ pp\\
Targeted misconception (1 item) & 65.9\% & 6.1\% & $-59.8$ pp\\
\end{tabular}
\end{ruledtabular}
\vspace{2pt}
\emph{Note.} The ``near-transfer $+$ far-transfer items'' row is a combined figure for the six items measuring high-symmetry point identification (3 items, near-transfer) and band--DOS reasoning (3 items, far-transfer); it is computed arithmetically from the formula $\text{Score}_{\text{remaining}} = (10\times\text{Score}_{\text{total}} - 4\times\text{Score}_{\text{core}})/6$. Separate near-transfer and far-transfer gains for these six items are not available because item-level response records were not retained. The combined figure of $+4.9$ pp does not imply that the two constituent constructs showed equal gains.
\rev{This figure is a derived quantity, not an independently measured one. Two caveats follow. (i) \emph{Rounding.} Because the total-score and core-item percentages are reported to three significant figures, the residual inherits their rounding and may deviate from the exact value by up to approximately $\pm0.2$ pp; the figure should therefore be read as ``about $+5$ pp'' rather than as a precise measurement. (ii) \emph{Partition assumption.} The identity holds only if the four core items, the six non-core items, and the total all refer to the same 10-item instrument with no overlap and no omissions. We confirm that the four core items cover two of the three constructs and that the six non-core items cover the remainder; however, we note explicitly that \emph{the targeted-misconception item is one of the four core items}. The core-item gain of $+27.1$ pp therefore includes the reversal of the targeted misconception, and the two rows of the table are not independent pieces of evidence.}
\end{table}

\begin{figure}[htbp]
  \centering
\center\includegraphics[width=12 cm,clip=true]{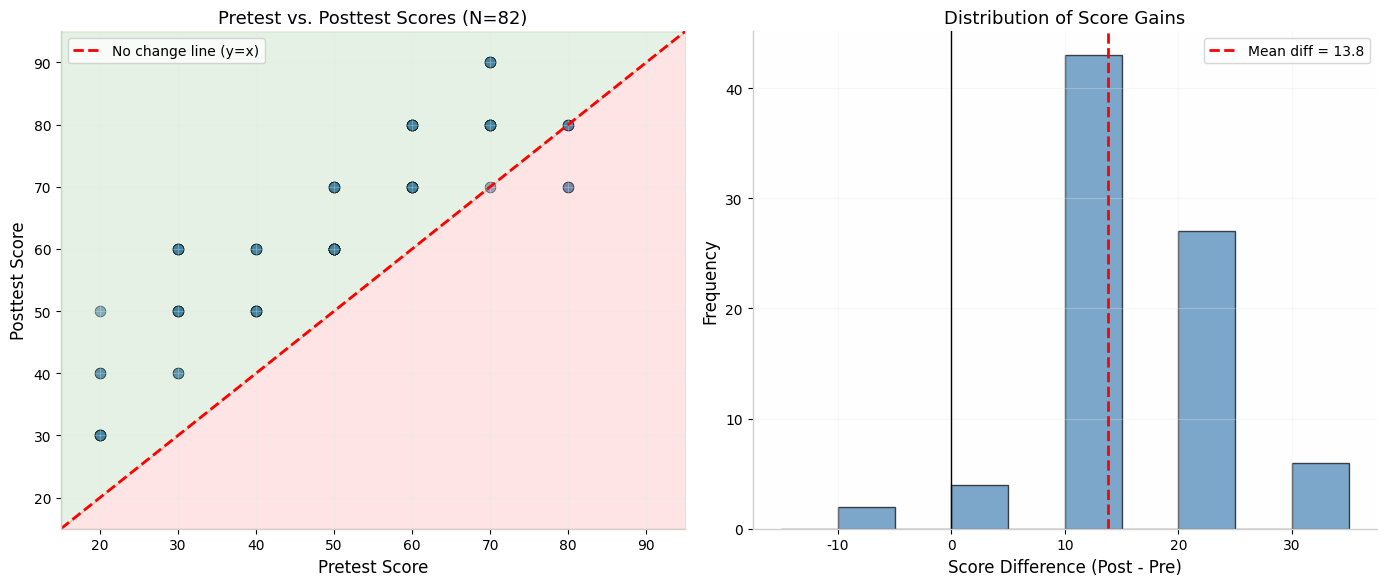}
\caption{\label{fig3}
Pretest versus posttest scores for the 82 students. Points above the y = x line indicate improvement}
\end{figure}

Cohorts 1 and 2 did not differ significantly on pretest or posttest, supporting data pooling, as show in the Fig.\ref{fig3}.

\rev{\emph{Artifactual explanations for the contrast.} Two features of this contrast bear on the most plausible artifactual explanations, and both point in the same direction. First, the six non-core items had a \emph{lower} pretest score (46.1\%) than the four core items (54.6\%). Because the non-core group starts farther from the ceiling, both regression to the mean and any general ``room to improve'' effect predict \emph{larger} gains for the non-core items; the observed ordering is the opposite ($+4.9$ vs.\ $+27.1$ pp). Second, uniform test--retest inflation predicts comparable gains across item groups, which is likewise not observed. We emphasize that this is an argument about the \emph{direction} of the discrepancy, not a statistical test: we cannot attach a confidence interval to the contrast because the item-level records were not retained (Sec.~IV.C), and we therefore do not claim that the numerical difference is statistically reliable. What the direction does establish is that the pattern is opposite to the two most common artifacts, which weakens---without eliminating---the artifactual reading.}

\emph{On interpreting the effect size.} In a paired pretest--posttest design, $d_z = t/\sqrt{n}$, and its magnitude is strongly influenced by the pretest--posttest correlation. In this study the pretest and posttest were highly correlated ($r=0.886$), which means that $d_z$ reflects not only the mean gain but also the consistency of individual gains. Readers should not equate $d_z=1.73$ directly with Cohen's $d=1.73$ from an independent-samples design.

As a complement, we report the common-language effect size: the probability that a randomly selected student's posttest score exceeds his or her pretest score is 0.93 (i.e., 93\% of students showed a gain, 4 showed no change, and 2 showed a slight decline). This index is not affected by the paired correlation and provides an additional interpretive dimension for the effect size.

The unusually large effect size may also reflect: the narrow alignment between instruction and assessment (4 of 10 items directly measure the core bottleneck), the absence of a ceiling constraint on the pretest, the single-institution and single-instructor context, and the fact that the parallel forms were judged rather than empirically equated for difficulty. Complete statistical details and examples of classroom feedback are provided in the Supplemental Material [31].

The purpose of reporting these results is not to provide evidence for the framework, but to show how the framework generates observable predictions in an actual classroom---namely, that an intervention identified as high-precision should show gains concentrated on the core-bottleneck items. In the present case, the four core-bottleneck items gained $+27.1$ pp, while the six near-transfer $+$ far-transfer items gained $+4.9$ pp in combination. This contrast is the differential-precision pattern referred to in Sec.~II.E. It should be read descriptively: the three item groups differ in content as well as in targeting, and item-level disaggregation of the six non-core items was not possible because item-level records were not retained. Despite the inability to disaggregate, the order-of-magnitude difference between the combined gain ($+4.9$ pp) and the core-bottleneck gain ($+27.1$ pp) still supports the qualitative interpretation of the pattern.

\section{DISCUSSION: MECHANISM, BOUNDARY CONDITIONS, AND LIMITATIONS}

\subsection{Patterns consistent with predictions}

The differential-precision pattern observed in this case---a $+27.1$ pp gain on the four core-bottleneck items versus $+4.9$ pp on the six near/far-transfer items---is consistent with the framework's prediction and inconsistent with uniform test--retest inflation, which would predict comparable gains on all items regardless of content. But as emphasized in Sec.~II.E, this pattern alone does not constitute a direct test of the precision mechanism, because the design of a high-precision intervention entails that some concentration is analytically expected, and because the three item groups differ in content as well as in targeting. The function of the case is to show how the framework generates observable predictions and to observe them in a concrete setting.

\subsection{Boundary conditions}

The framework's applicability is not unconditional. The precision advantage may diminish when: (i) the cognitive bottleneck is ill-defined or highly individualized; (ii) the representational tool introduces new extraneous load (e.g., a complex software interface); (iii) students lack the prior knowledge needed to interpret the visualization; (iv) precision approaches its upper bound, so that there is no further room to reduce extraneous load; (v) \emph{expertise reversal effect}: when students already possess the schema for the bottleneck transformation, a high-$P_r$ externalized representation may produce a redundancy effect, or even impede automatization [30]. The expertise reversal effect [30] directly corresponds to the framework's boundary conditions: the advantage of $P_r$ is moderated by the learner's prior knowledge. A high-precision externalization effective for novices may no longer be beneficial for experts. Dosage requirements may dominate when: (i) the learning goal involves skill development rather than conceptual change; (ii) multiple interconnected bottlenecks must be addressed simultaneously; (iii) long-term retention and transfer are the primary objectives; or (iv) dosage falls below the minimum effective processing threshold $D_{\min}$. As derived in Sec.~II.C, the precision advantage should vary with the representational nature of the bottleneck: the more the bottleneck depends on spatial--representational processing, the larger the marginal benefit of precision (Prediction 5). These boundary conditions themselves generate testable predictions---for example, the precision advantage should be smaller for bottlenecks whose difficulty is primarily procedural than for those whose difficulty is primarily representational.

\subsection{Limitations}

Several limitations constrain the strength of the empirical support for the framework proposed here. First, the illustrative case uses a one-group pretest--posttest design without a control or comparison condition, so the observed gains are reported as temporal associations rather than causal effects. Second, the sample comes from two consecutive cohorts at a single institution. Third, item-level response data were not retained, so internal-consistency statistics cannot be computed. Fourth, the unusually large effect size in the case should be interpreted in light of design features such as the narrow alignment between instruction and assessment, the single-institution context, and the fact that the parallel forms were not empirically equated for difficulty; the standardized effect size should not be treated as directly comparable to effect sizes from broader instructional studies. Fifth, the classroom feedback is descriptive and formative rather than a systematically coded dataset. Sixth, the differential-precision pattern is supported by the contrast between the core-bottleneck subscale ($+27.1$ pp) and the combined near/far-transfer items ($+4.9$ pp), rather than by a direct item-by-item comparison; item-level response records were not retained, so the six non-core items could not be disaggregated into their two constituent constructs of near-transfer (high-symmetry point identification) and far-transfer (band--DOS reasoning). Nevertheless, the order-of-magnitude difference between the combined gain ($+4.9$ pp) and the core-bottleneck gain ($+27.1$ pp) still supports the qualitative interpretation of the pattern. These limitations do not affect the theoretical contribution of the framework, but they do constrain the degree of empirical support the case can provide. Testing the framework requires future multi-condition, multi-institution studies.

\rev{\emph{A low-cost next step that does not require new student data.} The most immediate way to strengthen the empirical content of this work is to convert $P_r$ from a design judgment into a rated quantity. This requires no new student measurements: three to five solid-state physics instructors unfamiliar with the framework can independently rate, on the 5-point scale defined in Sec.~II.E, the extent to which each of the two representations (the interactive 3D visualization used here, and the conventional 2D blackboard cross-sections it replaced) eliminates the need for students to perform mental rotation. Reporting the resulting inter-rater agreement (e.g., ICC) would (i) provide validity evidence that the two representations differ in $P_r$, which is currently assumed rather than shown, and (ii) partially address the objection that ``2D versus 3D'' may not correspond to ``low versus high $P_r$'' in the intended sense. We identify this as the single most cost-effective extension of the present study, and we recommend it as a minimum requirement for any future comparative test built on this case.}

\section{CONCLUSION}

The main contribution of this paper is to propose a dosage--precision framework for instructional design, anchor it in cognitive load theory, and derive five falsifiable predictions (differential-precision pattern, conservation of time, dose--response curvature, mechanism transfer, and bottleneck-nature moderation). The framework's incremental contribution is not to rediscover cognitive load theory, nor to restate Carroll's time-needed model, but to decompose ``precision'' into separable design components, to show that only representational match acts on extraneous load, and to derive from this a quantifiable substitutability rate between precision and dosage. Under the formalization, the framework's core claim is compact: with $\eta(P_r)$ the representational efficiency, effective dosage $D_{\mathrm{eff}}=D P_s$, germane processing $G=D P_s\eta(P_r)$, and learning $E=g(G)$, the substitution rate along an iso-germane contour is $\mathrm{d}D/\mathrm{d}P_r=-\eta/\eta'$; the claim is bounded by a precision ceiling and a minimum dosage threshold, and the most falsifiable component is the predicted dependence of $\eta'(P_r)/\eta(P_r)$ on the representational nature of the bottleneck.

As an illustrative case, we reported a 15-minute three-dimensional visualization intervention in which patterns consistent with the framework's predictions appeared: gains were concentrated on the core-bottleneck items ($+27.1$ pp), with a substantially smaller combined gain on the near-transfer and far-transfer items ($+4.9$ pp). The case illustrates rather than tests the framework, because its single-condition design does not support comparative tests of P1, P2, P3, or Prediction 5.

Whether precision can compensate for limited dosage remains a falsifiable hypothesis for future controlled, multi-condition studies. The practical implication of the framework is that when class time is scarce, the most effective response is often not to cover more, but to find the precise point at which students fail---and to spend the available minutes there. The value of the framework lies in the research agenda it defines and in how it reformulates the problem faced by instructors in every discipline---how to allocate scarce instructional time---as a question about the interaction of two design variables.

\emph{Generalizability beyond solid-state physics.} The two foundational assumptions of the framework---identifiable cognitive bottlenecks and mechanism-matched representational tools---are domain-general, suggesting that its predictions may extend to any learning domain constrained by a spatial or representational bottleneck; this extension is itself the falsifiable Prediction 5.

\begin{acknowledgments}
This work was carried out at the National Supercomputing Center in Tianjin, and the calculations were performed on the Tianhe new-generation supercomputer.
\end{acknowledgments}


\section*{CONFLICT OF INTEREST}
The authors declare no competing interests.


\section*{DATA AVAILABILITY}
All instructional materials (VESTA session files, pre-computed DFT outputs, worksheet prompts) and both complete assessment forms are provided in the Supplemental Material. Anonymized total-score data are available from the corresponding author upon reasonable request. Item-level response data were not retained and cannot be made available.

\end{document}